\documentclass[12pt]{article}
\usepackage{amssymb,amsmath}
\usepackage{epsfig}

\newcommand{\tr}{\mathop{\mathrm{tr}}\nolimits}

\newcommand{\non}{\nonumber \\*}
\def\op{\oplus}
\def\ot{\otimes}
\def\P{{\cal P}}
\def\one{{\bf 1}}
\def\spin{{\bf S}}
\def\sc{\spin_l\cdot\spin_{l+1}}

\def\4N{${\cal N}=4$}

\begin{document}

\renewcommand{\thefootnote}{\fnsymbol{footnote}}
\setcounter{footnote}{1}

\begin{titlepage}
\begin{flushright}%\footnotesize\tt
%\\
ITEP-TH-22/04\\
UUITP-05/04\\
\end{flushright}
\vspace{.5cm}

\begin{center}
{\LARGE \bf New Integrable Structures} \\
\vskip 0.2cm
{\LARGE \bf in Large-N QCD}\\
\vspace{1.6cm}
{\large Gabriele Ferretti$^a$, Rainer Heise$^a$
and Konstantin Zarembo$^b$\footnote{Also at ITEP, Moscow, Russia}}
\vskip 0.8cm
{\large \it $^a$Institute for Theoretical Physics\\
G{\"o}teborg University and Chalmers University of Technology, \\
412 96 G{\"o}teborg, Sweden}
\vskip 0.5cm
{\large \it $^b$Department of Theoretical Physics\\
Uppsala University, 751 08 Uppsala Sweden  }
\end{center}

\vspace{0.5cm}
\begin{abstract}
We study the anomalous dimensions of single trace operators composed
of field strengths $F_{\mu\nu}$ in large-$N$ QCD.
The matrix of anomalous
dimensions is the Hamiltonian of a compact spin chain
with two spin one representations at each vertex corresponding to the selfdual
and anti-selfdual components of $F_{\mu\nu}$. Due to the special form of
the interaction it is possible to study separately renormalization of
purely selfdual components. In this sector the Hamiltonian is integrable
and can be exactly solved by Bethe ansatz.
Its continuum limit is described by the level two
$SU(2)$ WZW model.
\end{abstract}

\end{titlepage}

\setcounter{page}{1}
\renewcommand{\thefootnote}{\arabic{footnote}}
\setcounter{footnote}{0}

The anomalous dimensions of local operators are important physical
quantities which are indispensable
in describing logarithmic scaling violation in QCD
\cite{Georgi:1951sr,Gross:ju}
and which have many uses in QCD phenomenology.
There are however other theoretical reasons for the study of
anomalous dimensions.
We hope that the study of anomalous dimensions and operator mixing
will help in understanding the large-$N$
limit of QCD \cite{'tHooft:1973jz} and in elucidating
its relationship to string theory \cite{Polyakov}.

It is not clear at present what kind of string theory, if any,
describes the large-$N$ limit of QCD, but for its superconformal cousin,
\4N supersymmetric Yang-Mills (SYM) theory,
the answer is widely believed to be known. The SYM is exactly dual to type IIB
superstring theory in $AdS_5\times S^5$, according to the
conjectured AdS/CFT duality \cite{Maldacena:1998re}.
The scaling dimensions of local operators play an important role
in the AdS/CFT correspondence, since they are dual to the energies
of the string states \cite{Gubser:1998bc,Witten:1998qj}.
The spectrum of scaling dimensions in the field theory thus coincides with the
string spectrum. The latter is poorly known
because of the difficulties in quantizing strings in
$AdS_5\times S^5$, but the
semi-classical string states with large energies
\cite{Berenstein:2002jq,Gubser:2002tv} can be studied in much detail.
Such states are dual to operators with large
dimensions that often contain many constituent
fields \cite{Berenstein:2002jq}.
Computation of anomalous dimensions of such large operators is
a challenging problem, even at one loop, because of the operator mixing.
This problem drastically simplifies in the large-$N$ limit and reduces
to diagonalization of a Hamiltonian of a certain spin chain.

The spin chain that computes the complete one loop matrix of
anomalous dimensions in \4N SYM possesses the
amazing property of complete integrability
\cite{Minahan:2002ve,Beisert:2003tq,Beisert:2003yb,Beisert:2003ys}.
(Some results also extend to the first few higher loops for a restricted class
of operators.)
The spectrum of the spin chain can be computed
exactly by means of the Bethe ansatz \cite{Bethe:1931hc,Faddeev:1996iy}.
Comparison of the semi-classical Bethe states
with the classical strings in $AdS_5\times S^5$ has
led to many quantitative tests of the AdS/CFT correspondence
\cite{Frolov:2003qc,Beisert:2003xu,Frolov:2003tu,Beisert:2003ea,
Engquist:2003rn,Arutyunov:2003rg,
Tseytlin:2003ii,Serban:2004jf,Kristjansen:2004ei,Kazakov:2004qf}.
More recently, the equivalence between the spin chain and strings
was made more explicit at the level of effective actions
\cite{Kruczenski:2003gt,Kruczenski:2004kw,Hernandez:2004uw,Stefanski:2004cw}.
It turns out that the effective action of the
one loop spin chain can be interpreted as the string action
in a certain gauge,
and thus the mixing matrix of large operators in principle
carries some information about the
world-sheet dynamics of the dual string theory. We do not know the string dual
of QCD, but it should be possible to compute the mixing matrix of
local operators at one loop, which can shed new light on the string
description of the large-$N$ limit.

Integrable structures arise in many instances in QCD. They
were first found in the analysis of the parton evolution in the Regge
regime \cite{Lipatov:1993yb,Faddeev:1994zg}.
In a subsequent development, the mixing matrices of Wilson operators
with various quantum numbers were
identified with Hamiltonians of non-compact spin chains
\cite{Braun:1998id,Braun:1999te,Belitsky:1999bf}, which at large $N$
turn out to be
integrable for maximal-helicity operators.
%If not all of the
%parton helicities are aligned, the integrability is lost.

We will study a different set of operators built from the
gluon field strength
$F_{\mu\nu}=\partial _\mu A_\nu -\partial _\nu A_\mu +ig[A_\mu ,A_\nu ]$:
\begin{equation}\label{ops}
{\cal O}=\tr F_{\mu _{1}\nu _{1}}\ldots F_{\mu _{L}\nu _{L}}.
\end{equation}
Multiplicatively renormalizable operators are
linear combinations of those. In particular, we can contract indices
of different $F_{\mu \nu }$'s or contract indices with $\varepsilon^{\mu \nu \lambda \rho}$
thus allowing for pseudo-tensors characterized by the
presence of one ${\tilde F}_{\mu\nu}$.
More general operators may contain multiple traces, quarks and covariant
derivatives and in principle
mix with pure field strength operators.
However, the mixing with quark and multi-trace
operators is suppressed by powers of $1/N$.
The mixing with operators that contain derivatives
is not suppressed, but we shall explain below that the mixing with derivative operators
starts at two loops.
Hence, pure field strength operators form
a closed sector at one loop.

The mixing matrix is defined as a logarithmic
derivative of the renormalization factor in the UV
cutoff\footnote{We use the conventions of \cite{FTbook}. In many papers,
the mixing matrix is defined with the
opposite sign, and sometimes the definition of operators includes
factors of the QCD coupling
which shifts the anomalous dimensions by the beta-function.}:
$\Gamma=Z^{-1}\cdot dZ/d\ln\Lambda$.
The renormalization factor is defined by the requirement that
multiplication by $Z$ makes correlation functions of composite operators
${\cal O}$ finite:
${\cal O}_{\rm ren}^{\,\boldsymbol{\mu}}=
Z^{\,\boldsymbol{\mu}}_{\hphantom{\,\pmb{\mu}}\boldsymbol{\nu}}
{\cal O}^{\,\boldsymbol{\nu}}$.
Here $\boldsymbol{\mu}$ and $\boldsymbol{\nu}$ are multi-indices
that parameterize all possible
operators of the form (\ref{ops}).
The eigenvectors of the mixing matrix are
 multiplicatively renormalizable operators whose
anomalous dimensions are given by the eigenvalues.
The size of the mixing matrix rapidly grows
with $L$ and the resolution of the operator mixing becomes a more and more
complicated problem.

\begin{figure}[ht]
%\hspace*{2cm}
%\epsfysize=11cm
\centerline{\epsfig{file=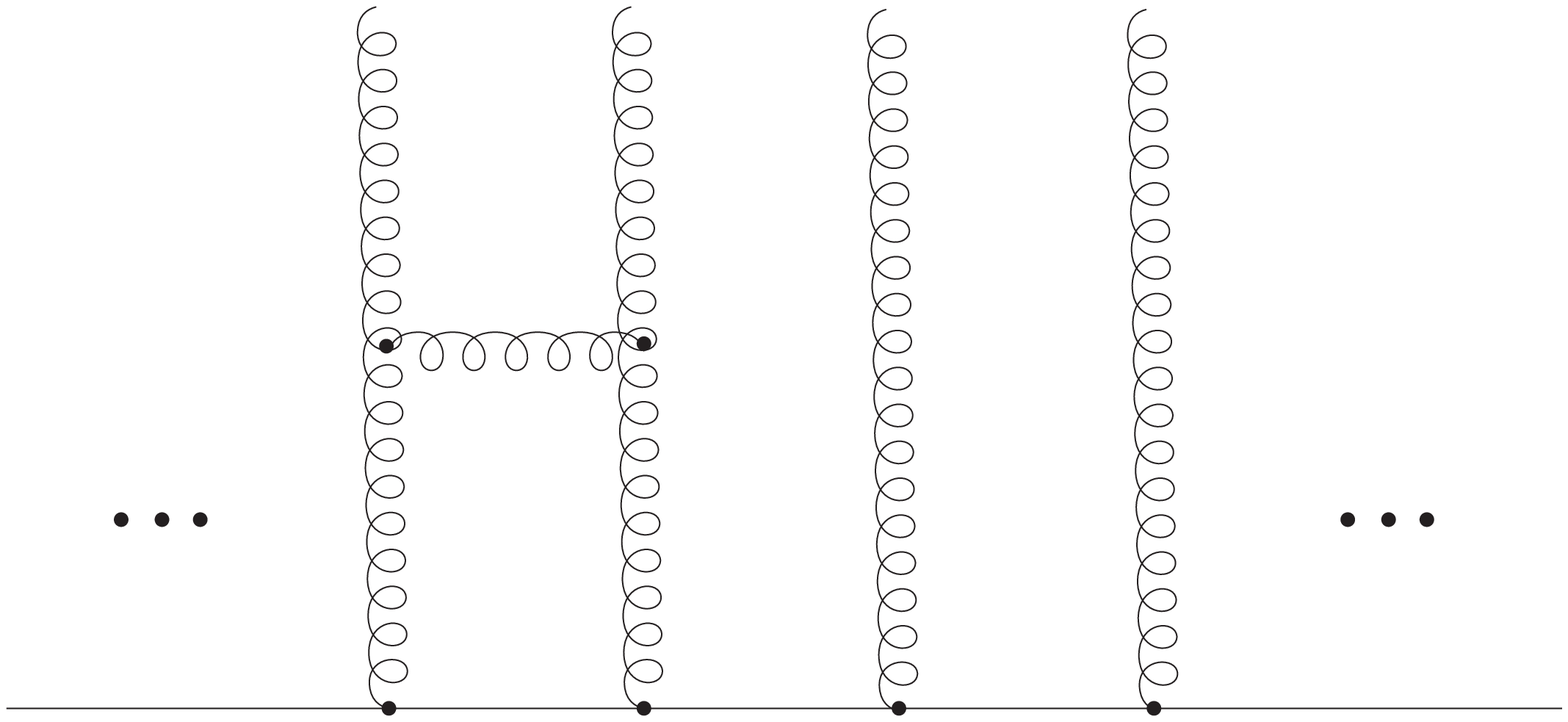,width=60mm}}
\caption[x]{\small One of the diagrams that contribute to the
mixing matrix at one loop.
The operator is depicted by a horizontal bar with gluon vertices ordered by
an overall trace over the color indices.}
\label{figone}
\end{figure}

Fortunately, a useful reformulation of this problem
drastically simplifies it in the large-$N$ limit.
The idea is to identify the operators (\ref{ops}) and their linear
combinations with the states in the Hilbert space of a periodic spin
chain with $L$ sites, one site for
each $F_{\mu_{l}\nu_{l}}$. The sites are naturally ordered by the trace over
the color indices. The mixing matrix is a Hermitian operator in
this Hilbert space and thus can be regarded as a Hamiltonian
of the spin chain.
The mixing matrix is determined by the divergent pieces of the
diagrams one of which is shown in Fig.~\ref{figone}.
In more general operators containing
covariant derivatives, $D_{\lambda_1}\cdots
D_{\lambda _n}F_{\mu _{l}\nu _{l}} $ should be identified with a
single site of the lattice (as in \cite{Braun:1998id,Braun:1999te,Belitsky:1999bf}),
which is clear from the structure of the diagrams
($D_{\lambda_1}\cdots
D_{\lambda _n}F_{\mu _{l}\nu _{l}} $ emits a single gluon
to the lowest order in perturbation theory).

It is also clear that the number of sites can diminish but cannot grow
at one loop, because a divergent diagram with $L+1$ external legs contains at least
three three-gluon vertices or one four-gluon and one three-gluon vertex
and consequently appears at $O(g^3)$.

A careful analysis \cite{Kluberg-Stern:rs} shows that
operators with derivatives, containing the equations of motion,
may appear at $O(g^2)$ in counterterms
for pure field strength operators, all other mixing being $O(g^3)$.
Thus, the one-loop
mixing matrix is block upper triangular (block diagonal if one
neglects operators vanishing by the equations of motion)
and the anomalous dimensions
of pure field strength operators are entirely determined by mixing among
themselves.
The other crucial property of the one-loop mixing matrix is
that is has only nearest-neighbor interactions at large $N$:
\begin{equation}\label{defgamma}
\Gamma = \frac{g^2N}{48\pi ^2}\sum_{l=1}^L
H_{l,l+1},
\end{equation}
where $H_{l,l+1}$ acts on the two adjacent sites.
Non-local interactions correspond
to non-planar diagrams and are suppressed at large $N$.

The two-body Hamiltonian $H_{l,l+1}$ acts on two $F_{\mu\nu}$'s and thus is
a matrix with eight Lorentz indices, which in principle can be computed
from the diagrams. Fortunately,
known anomalous dimensions of several particular operators
make this calculation unnecessary. The
Lorentz invariance severely restricts the structure of the mixing matrix
and significantly reduces the number of independent matrix elements. The field
strength transforms in the reducible representation $(1,0)\op(0,1)$
of the Lorentz group.
The irreducible components are self-dual and anti-self-dual
parts of $F_{\mu \nu }$:
\begin{equation}
F_{\mu \nu }=\sigma _{\mu\nu}^{\hphantom{\mu\nu}\alpha \beta }f_{\alpha \beta }
+\bar{\sigma} _{\mu\nu}^{\hphantom{\mu\nu}\dot{\alpha}\dot{\beta} }
\bar{f}_{\dot{\alpha}\dot{\beta} },
\end{equation}
where $\sigma _{\mu\nu}=i\sigma _2(\sigma _{\mu}\bar{\sigma}_{\nu}-
\sigma _{\nu}\bar{\sigma}_{\mu})/4 $,
 $\bar{\sigma} _{\mu\nu}=-i(\bar{\sigma} _{\mu}{\sigma}_{\nu }
-\bar{\sigma} _{\nu}{\sigma}_{\mu })\sigma _2/4 $,
$\sigma _\mu =(1,\boldsymbol{\sigma })$,
$\bar{\sigma }_\mu =(1,-\boldsymbol{\sigma })$.

The chiral pieces of the field strength $f_{\alpha \beta }$ and
$\bar{f}_{\dot{\alpha}\dot{\beta} }$ transform in the spin one representations
of $SU_L(2)$ and $SU_R(2)$. It is convenient
to introduce the vectors
$f_A=(\sigma_2\sigma_A) ^{\alpha \beta }f_{\alpha \beta }$ and
$\bar{f}_{\dot{A}}=(\sigma_2\sigma_{\dot{A}}) ^{\dot{\alpha }\dot{\beta }}
\bar{f}_{\dot{\alpha}\dot{\beta} }$, where $A,\; \dot{A} = 1,2,3$
and  $\sigma _A$, $\sigma _{\dot{A}}$ are ordinary Pauli matrices.
Since the Hamiltonian $H_{l,l+1}$ commutes with Lorentz transformations,
it takes constant values on irreducible representations that appear
in the tensor product
\begin{eqnarray}\label{tp}
[(1,0)\op(0,1)]\ot[(1,0)\op(0,1)]
&=&(2,0)\op(0,2)\op(1,0)\op(0,1)\op(0,0)^+
\non
&&\op(0,0)^- \op(1,1)^+ \op(1,1)^-,
\end{eqnarray}
where the superscripts denote parity.
The anomalous dimensions of representations related by
parity should be the same, so the Hamiltonian
contains six independent structures:
\begin{equation} \label{gr}
H=a(\P_{(2,0)}+\P_{(0,2)})+b(\P_{(1,0)}+\P_{(0,1)})+c\P_{(0,0)^+}+d\P_{(0,0)^-}
+e\P_{(1,1)^+}+f\P_{(1,1)^-},
\end{equation}
where $\P_R$ are projectors on  irreducible representations.
The coefficients $a$--$f$
can be fixed by comparing the energies of the eigenstates for short chains
$(L=2,3,4)$
with known anomalous dimensions.

We start with dimension four operators that correspond to the spin chain
with two sites connected by two links. The Hamiltonian (\ref{defgamma})
acts on each of the two links in the same way.
Because of the cyclicity of trace, only symmetric representations in
(\ref{tp}) survive.
The simplest operator is the energy-momentum tensor $T_{\mu\nu}$
which belongs to $(1,1)^+$ in (\ref{tp}). Since
the energy-momentum tensor has zero anomalous dimension, $e=0$.

Next, the action and the topological density are RG invariant
to one loop when multiplied by $g^2$ \cite{Kluberg-Stern:rs,Morozov:ef}.
The anomalous dimensions of the canonically normalized operators
are thus equal to the beta function:
\begin{equation}\label{ad1}
\tr F_{\mu\nu} F_{\mu\nu}, \;\;\; \tr F_{\mu\nu} \tilde{F}_{\mu\nu}:~~~~~
\gamma=-\frac{11g^2N}{24\pi^2}\,.
\end{equation}
Acting by (\ref{defgamma}), (\ref{gr})
on the scalar and on the pseudoscalar we obtain the anomalous dimensions
$2cg^2N/48\pi^2$ and $2dg^2N/48\pi^2$ respectively, which implies that $c=d=-11$.

The anomalous dimension of the $(2,0)+(0,2)$ operator was computed in
\cite{Robertson:1990bf}:
\begin{equation}\label{ad2}
\tr F_{\mu\nu} F_{\lambda \rho }-{\rm index~contractions}:~~~~~
\gamma =\frac{7g^2N}{24\pi^2}\,.
\end{equation}
Comparing to the eigenvalue of the spin chain Hamiltonian, we find: $a=7$.

The anomalous dimensions of the unique scalar and pseudoscalar operators
of dimension six are also known
\cite{Morozov:ef}:
\begin{equation}\label{ad3}
\tr F_{\mu\nu} F_{\nu\lambda } F_{\lambda \mu }, \;\;\;
\tr \tilde{F}_{\mu\nu} F_{\nu\lambda } F_{\lambda \mu }:
~~~~~\gamma =\frac{g^2N}{16\pi^2}\,.
\end{equation}
The calculation of the eigenvalues in the spin chain is no more tricky than for
dimension four states.
The chain now contains three sites. The Hilbert space consists of
three copies of $(1,0)\op(0,1)$. In order to get a scalar state,
we should keep
only $(1,0)$ and $(0,1)$ in the tensor product of basic representations, since only
$(1,0)$ and $(0,1)$ can produce a scalar when tensored with the remaining
$(1,0)\op(0,1)$ on the third site. Hence, only the projectors
$\P_{(1,0)}$ and $\P_{(0,1)}$ in the Hamiltonian contribute to the
energy. Thus, we get  $3bg^2N/48\pi^2$ for the anomalous dimension.
Consequently, $b=1$.

To compute the last factor $f$ we must consider operators of dimension eight.
There are four scalars at this level:
\begin{equation}
\tr F_{\mu\nu}F_{\mu\nu}F_{\rho\lambda}F_{\rho\lambda},\;
\tr F_{\mu\nu}F_{\rho\lambda}F_{\mu\nu}F_{\rho\lambda},\;
\tr F_{\mu\nu}F_{\nu\rho}F_{\rho\lambda}F_{\lambda\mu},\;
\tr F_{\mu\nu}F_{\nu\lambda}F_{\mu\rho}F_{\lambda\rho}
\end{equation}
and their mixing matrix can be found in \cite{Gracey:2002rf}:
\begin{equation}
     \Gamma = \frac{g^2 N}{48\pi^2}
     \begin{pmatrix}-17 & -2 & -18    & -14 \cr
             -2 & 8 & -20 & -12    \cr
             -11&    0 & 2  & 2   \cr
              -1&   -6 &  8  & -16  \end{pmatrix}. \label{gamma8}
\end{equation}
It is then a straightforward but slightly tedious exercise to show that
this results implies $f = 3$.

We have independently checked all the above results.
Also, the mixing matrix (\ref{gamma8}), in addition to fixing
$f=3$, yields several algebraic relations among the remaining
coefficients in the Hamiltonian all of which are satisfied by their
numerical values computed earlier.

To summarize, the spin chain Hamiltonian is
\begin{equation}
H=7(\P_{(2,0)}+\P_{(0,2)})+\P_{(1,0)}+\P_{(0,1)}-11(\P_{(0,0)^+}+\P_{(0,0)^-})
+3\P_{(1,1)^-}.
\label{spinchain}
\end{equation}
An important point to notice here is the degeneracy of the
scalar representations of opposite parity.
Because of this degeneracy, definite-parity projectors can be
replaced by projectors on the left and right scalars ($f_Af_A$ and
$\bar{f}_{\dot{A}}\bar{f}_{\dot{A}}$). As a result,
only the last term in the Hamiltonian
mixes  $f_A$ with $\bar{f}_{\dot{A}}$ and even this term conserves the number
of $f_A$'s (and of $\bar{f}_{\dot{A}}$'s). Hence, the number of
self-dual components of the field strength in an operator is preserved by
renormalization.
This follows from a ``chiral'' type of symmetry rotating
$f_A$ and $\bar{f}_{\dot{A}}$ by
opposite phases. The Lagrangian is not invariant under this transformation
but its variation is proportional to the topological density that
integrates to zero in the absence of instantons. This symmetry is
responsible for the closure of the chiral sector
$\tr(f_{A_1}\dots f_{A_L})$.
Such structure of the mixing matrix produces
parity degeneracies for
operators of higher dimension as well. For instance,
parity-odd and parity-even operators in $(L,0)+(0,L)$
and $(L,0)-(0,L)$ are degenerate.
The first example of such parity degeneracy is (\ref{ad3}).

The spin system (\ref{spinchain}) can be represented as a spin
ladder.
Let us put all six states at each site into a $3\times 2$ matrix $(f_A,\bar{f}_{\dot{A}})$.
One can define two sets of spin operators,
which act on this matrix vertically and horizontally:
\begin{equation}
(S^i\,f)_A=i\varepsilon_{iAB}f_B,\qquad (S^i\,\bar{f})_{\dot{A}}=i\varepsilon_{i\dot{A}\dot{B}}\bar{f}_{\dot{B}},
\end{equation}
\begin{eqnarray}
&&\tau ^3\,f=f,\qquad \tau ^+\,f=0,\qquad \tau ^-\,f=\bar{f},
\nonumber  \\*
&&\tau ^3\,\bar{f}=-\bar{f},\qquad \tau ^+\,\bar{f}=f,\qquad \tau ^-\,\bar{f}=0.
\end{eqnarray}
The selfdual and anti-selfdual components of $F_{\mu\nu}$ can be visualized as
spins sitting on two parallel spin chains.
The mixing matrix can be written as
\begin{eqnarray}
\Gamma &=&\frac{g^2N}{48\pi ^2}\,\sum_{l=1}^{L}\left\{
\frac{1}{2}\left(1+\tau ^3_l\tau ^3_{l+1}\right)
\left[7+3\sc-3(\sc)^2\right]
\right.
\nonumber  \\*
&&\left.
+\frac{3}{4}\left(1-\boldsymbol{\tau}_l\cdot\boldsymbol{\tau}_{l+1}\right)
\right\} .
\label{mmattix}
\end{eqnarray}

The Hamiltonian possesses an $SU(2)\times SU(2)\times U(1)$
symmetry. The total spin
\begin{equation}
\spin=\sum_{l=1}^{L}\spin_l
\end{equation}
generates spatial rotations, while Lorentz boosts are generated by
\begin{equation}
{\bf K}=\sum_{l=1}^{L}\spin_l\tau ^3_l.
\end{equation}
The chiral components of the field strength carry charges $\pm 1$
under the extra $U(1)$ symmetry:
\begin{equation}
\chi =\sum_{l=1}^{L}\tau ^3_l.
\end{equation}
Because of this symmetry the number of self-dual and anti-self-dual
components of the field strength in an operator are independently
conserved, and we can consider sectors with a fixed number of
$f_A$'s and $\bar{f}_{\dot{A}}$'s.

The simplest and the most interesting sector is composed of
operators that contain only the self-dual field strength $f_A$. The
mixing matrix (\ref{mmattix}) then reduces to
\begin{equation}
\Gamma =\frac{g^2N}{48\pi ^2}\sum_{l=1}^L
\left[7+3\sc-3(\sc)^2\right].
\label{intspinchain}
\end{equation}
The mixing matrix in this sector is a Hamiltonian
of a spin one $SU_L(2)$ spin chain.
Alternatively, (\ref{spinchain}), restricted to the chiral sector, can be expressed in terms of permutation and trace operators
acting on three-dimensional vectors as
\begin{eqnarray}
P\,v\ot u&=&u\ot v,
\non
K\,v\ot u&=&(v,u)\one,
\end{eqnarray}
These operators are $SU(2)$ invariant and can be expressed in terms
of projectors on irreducible representations:
\begin{equation}
P=\P_0-\P_1+\P_2,~~~~~K=3\P_0.
\end{equation}
 In terms of the permutation
and trace operators, the mixing matrix of chiral operators is
\begin{equation}\label{PK}
\Gamma =\frac{g^2N}{48\pi ^2}\sum_{l=1}^L
\left(4+3P_{l,l+1}-6K_{l,l+1}\right).
\end{equation}

Because the spins repel, the vacuum of the system is anti-ferromagnetic,
which means that the  lowest anomalous dimension is negative.
Remarkably, the spin chain (\ref{intspinchain}) is integrable
\cite{Zamolodchikov:ku,Kulish:gi,Resh2,Faddeev:1996iy}
and the Hamiltonian can be diagonalized by Bethe
ansatz \cite{ttjan,Babujian:1982ib}!
The Bethe equations constitute a set of algebraic equations
for rapidities of elementary excitations on the lattice:
\begin{equation}
\left(\frac{\lambda _k+i}{\lambda _k-i}\right)^L=
\prod_{j\neq k}\frac{\lambda _k-\lambda _j+i}{\lambda _k-\lambda _j-i}\,,
\end{equation}
The Bethe equations should be supplemented by  the zero-momentum condition:
\begin{equation}
\prod_k\frac{\lambda _k+i}{\lambda _k-i}=1.
\end{equation}
This condition reflects the cyclicity of the trace in the
operators.
All possible solutions of Bethe equations yield the spectrum of the anomalous
dimensions, computed for a given solution from the formula
\begin{equation}
\gamma =\frac{g^2N}{48\pi ^2}\left(7 L -\sum_k\frac{12}{\lambda _k^2+1}\right).
\end{equation}

Some sample anomalous dimensions can be readily calculated.
The state
with the largest possible spin corresponds to the maximally symmetric
operator
\begin{equation}
{\cal O}_{\Omega}=\tr f_{(A_1}\ldots f_{A_L)}-{\rm index~contractions}.
\end{equation}
This operator corresponds to the pseudo-vacuum of the spin chain. Its
anomalous dimension is large and positive:
\begin{equation}
\gamma _{\Omega }=\frac{7g^2N}{48\pi ^2}\,L.
\end{equation}
At $L=3$,
this formula reproduces the known anomalous dimension of
twist three dimension six gluon operator \cite{Belitsky:1999bf}.
The real vacuum is a Lorentz scalar and has negative anomalous dimension (as
an effect of the asymptotic freedom).
In the thermodynamic limit of large $L$ \cite{ttjan,Babujian:1982ib}:
\begin{equation}
\gamma _0=-\frac{5g^2N}{48\pi^2}\,L +O(L^0).
\end{equation}

The known string representation of the superconformal Yang-Mills theory
can be compared to the spin chain rather directly. The
classical spin system obtained by
taking the continuum limit of the relevant lattice model (which is
of course different from the one considered here) agrees with that
describing the semiclassical rotating string in the bulk spacetime
\cite{Kruczenski:2003gt,Kruczenski:2004kw,Hernandez:2004uw,Stefanski:2004cw}.
This observation opens up the possibility of reconstructing the
string description %of a certain gauge theory
by analyzing the large-$N$
renormalization of operators that consist of a large number of elementary
fields. Needless to say, this map is far from complete
since on the gauge theory side we have limited information on higher loop
effects and %even conceptually
the mapping becomes more involved as these effects are taken into account.
Still, we believe it is of interest to consider the continuum limit of the
spin chain we have obtained, hoping that it will provide some clue on
the string description.

The continuum limit of
the anti-ferromagnetic spin one chain is a relativistic theory  in the
integrable case, because
the dispersion relation
of low-lying excitations is linear \cite{ttjan,Faddeev:1981ip,Faddeev:1996iy}.
The low-energy effective theory is the conformal
$SU(2)$  WZW model at level two \cite{Affleck:wb, Affleck:1987ch}.
This model has central charge $c = 3/2$ and can be described
by a triplet of free fermions:
%These facts can be easily understood by looking at low-lying excitations around
%the anti-ferromagnetic vacuum which form a triplet of fermions with
%the linear dispersion relation \cite{ttjan,Faddeev:1981ip,Faddeev:1996iy}
%and are thus described by the relativistic action in the continuum limit:
\begin{equation}
S=\int d^2\xi \,\sum_{a=1}^{3}\left(\chi _L^a\partial _+\chi _L^a
+\chi _R^a\partial _-\chi _R^a\right).
\end{equation}
The fermion currents $J_L^a =i \varepsilon^{abc}  \chi_L^{b}\chi_L^c/2$
(and similarly for $J_R$) obey a $SU(2)$ Kac-Moody algebra at level two:
\begin{equation}
{[J_L^a(x_-), J_L^b(y_-)]} = i \epsilon^{abc} J_L^c(x_-)\delta(x_--y_-) +
      2\,\frac{i}{4\pi}\delta^{ab}\delta^\prime(x_--y_-).
\end{equation}
The current algebra can be bosonized in the standard way \cite{Witten:ar}
by introducing the chiral field $g(x) \in SU(2)$
with the WZW action
\begin{equation}
    S=\frac{2}{16\pi}\int d^2\xi  \;
        \tr \partial_\alpha g\partial^\alpha g^\dagger+
      \frac{2}{24\pi}\int d^3\xi  \epsilon^{\alpha\beta\gamma}
        \tr g^\dagger\partial_\alpha g g^\dagger\partial_\beta g
            g^\dagger\partial_\gamma g.
\end{equation}
Alternatively, one can trade two out of three fermions
for a compact boson. In this representation, the model
is manifestly supersymmetric.

We are thus led to believe that some of the dynamics of
the purely selfdual sector
of QCD in the large $N$ and $L$ limit  is captured by a
WZW model, although higher loop corrections and the inclusion
of derivatives need to be carefully analyzed (work in progress).
We hope that this results will provide further clues on the construction
of the elusive QCD string.

{\it Acknowledgments}: We would like to thank H.~Johannesson and J.~Minahan
for discussions and A.~Gorsky for pointing out reference \cite{Morozov:ef} to us.
The research of G.F. is supported by the Swedish Research Council
(Vetenskapsr\aa det) contract 622-2003-1124.
The work of K.Z was supported in part by the Swedish Research Council
under contract 621-2002-3920 and by G{\"o}ran Gustafsson Foundation.

\end{document}